\documentclass{an}
\usepackage{graphicx}
\usepackage{times}
\Volume{0}               
\Year{2004}              
\Pagespan{000}{000}      

\begin{document}
\lhead[\thepage]{P. Petit et al.~: Differential rotation of stars}
\rhead[Astron. Nachr./AN~{\bf 325}, No. \volume\ (\yearofpublic)]{\thepage}
\headnote{Astron. Nachr./AN {\bf 325}, No. \volume, \pages\ (\yearofpublic) /
{\bf DOI} 10.1002/asna.\yearofpublic1XXXX}

\title{Differential rotation of cool active stars}

\author{P. Petit$^1$ \and J.-F. Donati$^2$ \and A. Collier Cameron$^3$}
\institute{Centro de Astrofisica da Universidade do Porto, rua das Estrelas, 4150-762 Porto, Portugal \and Observatoire Midi-Pyr\'en\'ees, 14 avenue Edouard Belin, 31400 Toulouse, France \and School of Physics and Astronomy, Univ. of St Andrews, Saint Andrews, Scotland KY16 9SS}
\date{Received {date}; 
accepted {date};
published online {date}} 

\abstract{
The surface differential rotation of active solar-type stars can be investigated by means of Doppler and Zeeman-Doppler Imaging, both techniques enabling one to estimate the short-term temporal evolution of photospheric structures (cools spots or magnetic regions). After describing the main modeling tools recently developed to guarantee a precise analysis of differential rotation in this framework, we detail the main results obtained for a small number of active G and K fast rotating stars. We evoke in particular some preliminary trends that can be derived from this sample, bearing the promise that major advances in this field will be achieved with to the new generation of spectropolarimeters (ESPaDOnS/CFHT, NARVAL/TBL).     
}
\correspondence{petit@astro.up.pt}

\maketitle

\section{Introduction~: differential rotation and dynamo activity}

Differential rotation is one of the basic ingredients invoked to explain the generation of the solar magnetic field, through its ability to transform a large-scale poloidal field into a stronger toroidal component. However, much details of this general principle are still poorly understood, and a major aim for stellar differential rotation measurements is to evaluate, in the stellar parameter space, how stellar activity is connected to differential rotation.

The photospheric magnetic field of several fast rotating stars has been mapped by means of Zeeman-Doppler Imaging (Donati \& Brown 1997), providing a wealth of informations about the dynamo processes at work in these extremely active objects. The most significant and puzzling result obtained in this framework is the repeated detection of large magnetic regions in which the orientation of the field is mainly azimuthal. This kind of structure is not observed on the Sun, and classical dynamo models suggest that such large-scale azimuthal components of the field may be closely linked to the effect of differential rotation. An analysis of differential rotation may therefore bring a pivotal element for understanding how the magnetic field of these objects is generated. 

Zeeman-Doppler Imaging is able to perform this task by detecting the short-term evolution of surface structures, using cool spots or magnetic regions as tracers of the surface shear. In this article, we first review the main modeling tools now available to investigate the differential rotation of late-type fast rotators. We then detail the main results obtained in this field and propose some promising directions in which this work may go ahead with the new generation of high resolution stellar spectropolarimeters.

\section{Extracting the differential rotation signal}

\subsection{Basic idea and required performances}

The very first idea that comes to mind for detecting a relative motion of active regions at the surface of a star is to compare successive images of the same object secured a few days apart, thus reproducing the patient work of Solar observers since the beginning of Solar activity systematic monitoring in the XVII$\rm^{th}$ century. Assuming a star with an intensity of the surface shear similar to that of the Sun, the maximum relative longitudinal motion of a pair of spots (one located near the pole and the second one close to the equator) does not exceed $3^\circ$ per day. In practice, the situation can be much more restrictive, owing for instance to a weaker surface shear or to a spot distribution spanning a smaller range of latitudes. As a consequence, the differential rotation signal is likely to be very tiny and comparisons of Doppler images have to face three important drawbacks. 

First, the time-series of spectra used to reconstruct each stellar map must be secured on a time-span shorter than any significant evolution of the surface, including differential rotation itself, so that every images can be considered as snapshots of the star. This constraint can only be fulfilled for ultra-fast rotators, with rotation periods typically shorter than one day. Unfortunately, this is not the case for stars rotating with a period of a few days, for which several weeks are usually necessary to achieve a dense phase sampling in the case of single-site observations. In this last case, reducing the time-span implies a sparse phase sampling of reconstructed images, with the well-documented effect of leading to inaccurate location of spots (errors in spot location being potentially larger than the differential rotation signal itself, Petit et al. 2002). 

Secondly, the successive maps used for comparison must be spaced by a time interval shorter than the lifetime of surface structures used as tracers of the shear. If this condition is not respected, new-born active regions can be mistaken for older ones that have vanished in the meantime, an aliasing problem leading to inadequate evaluations of the differential rotation. While photospheres of single stars evolve on timescales as short as a couple of weeks (Barnes et al. 1998, Petit et al. 2003b), components of close binary systems are more stables, with typical evolution of active regions of the order of one month (Petit et al. 2003a).

Finally, the differential rotation signal is very weak in many cases and can therefore be unambiguously detected only in optimal observing conditions. In this context, a low noise level is critical, so that the use of multi-lines techniques like the Least-Squares Deconvolution proposed by Donati et al. (1997) is almost compulsory.

Some extensive works were based on comparisons of cool spot maps (see, e.g., Vogt et al. 1999, Strassmeier \& Bartus 2000, Rice \& Strassmeier 1996), though it is retrospectively clear that these studies do not respect the whole set of conditions detailed above. Some firm results were however obtained in this framework, in the very suitable cases of the young dwarfs AB~Dor and PZ~Tel (Donati \& Cameron 1997, Donati et al. 1999, Barnes et al. 2000). The development of specific tools was however required in order to obtain more accurate estimates of the shear and more reliable error bars, as well as to enlarge the potential number of targets adapted to a study of differential rotation.
 
\subsection{Direct spot tracking}

The first of these tools, although not truly speaking Doppler imaging, is nevertheless based on similar principles and makes use of the same kind of data sets, i.e. densely time sampled series of high resolution spectra.

This method is directly inspired by the tomographic techniques developed to track prominences around solar-type stars (Cameron \& Robinson 1989), then adapted to detect starlight reflected by extra-solar planets (Cameron et al. 2002). A time-series of high S/N line profiles, extracted by means of Least-Squares Deconvolution, is used for a direct tracking of star-spot signatures inside the spectra themselves. A matched-filter analysis is employed to determine the area, velocity amplitude and rotation period of individual spots, allowing to determine the rotation period of active regions as a function of stellar latitude (without any assumption on the underlying rotation law) and offering the original and potentially fruitful opportunity to select the tracers according to their area (Cameron et al. 2002, Cameron \& Donati 2002).

\subsection{Parametric imaging}

\begin{figure}[t]
\resizebox{\hsize}{!}
{\includegraphics[angle=270]{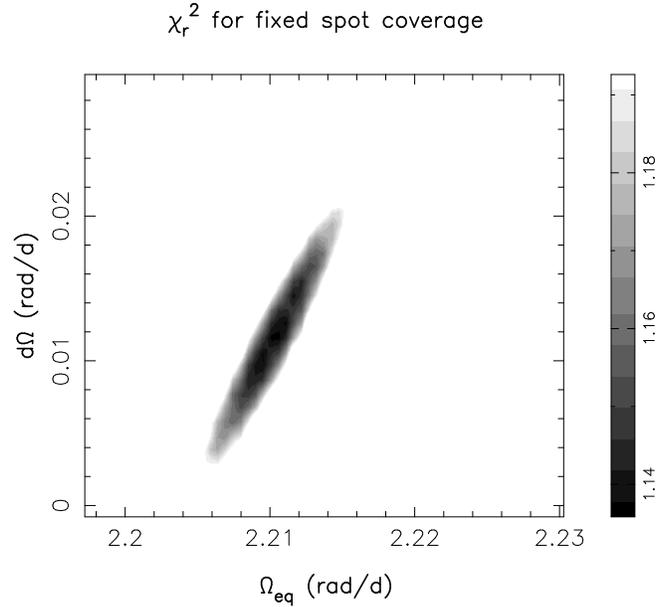}}
\caption{reduced $\chi^2$ map in the differential rotation parameter plane, obtained for HR~1099 from polarized spectral data, using a parametric imaging method. $\Omega_{\rm eq}$ is the rotation rate of the equator and $\rm d\Omega$ the difference in rotation rate between the pole and the equator. The 1$\sigma$ limit on the two parameters, considered separately, corresponds to the black region.}
\label{hr1099}
\end{figure}

Another method, based on Zeeman-Doppler Imaging, consists in assuming that the differential rotation can be treated as any other stellar parameter into the imaging process. According to the principle of maximum entropy reconstruction, the most likely set of parameters of the differential rotation law is the one that minimizes the information content of the image (at a fixed $\chi^2$). Equivalently, the most likely differential rotation law is the one that minimizes the $\chi^2$ of the reconstructed map at a fixed information content, this last option ensuring an easy access to error bars. Assuming a solar-like rotation law of the form~: $\Omega(l) = \Omega_{\rm eq} - {\rm d}\Omega \sin^2 l$ (where $\Omega(l)$ is the rotation rate at latitude $l$, $\Omega_{\rm eq}$ the rotation rate of the equator and $\rm d\Omega$ the difference in rotation rate between the pole and the equator), several maps are reconstructed for different values of the two parameters, in order to determine the pair that minimizes the $\chi^2$ of the reconstructed spectra (Fig. \ref{hr1099}).

Any type of differential rotation law can be implemented this way, but a simple solar-like law appears to be a good compromise, since only two parameters have to be optimized. Even for a star like AB~Dor, for which the differential rotation signal has been studied at a high precision level with direct spot tracking (Cameron \& Donati 2002), such a simple law provides a very good fit to the data.

This method, first employed by Donati et al. (2000), was tested by Petit et al. (2002) through a series of numerical simulations. It was demonstrated in particular that its performances can be satisfactory for a large range of stellar rotation periods. Comparisons with direct spot tracking measurements have demonstrated that differences between both methods remain acceptable. 

Cool spots and magnetic structures can be independently investigated by parametric imaging, a very interesting opportunity since for fast rotating stars the magnetic topology usually offers tracers densely distributed over the surface, while in some cases cool spots are mostly concentrated close to the pole, with few tracers available at low latitude. Moreover, potential aliasing between neighbouring magnetic regions is less problematic than for cool spots, because of the additional information provided by the polarity of their magnetic field.

\section{Application to a sample of stars}

Differential rotation was successfully estimated on a small sample of active stars of spectral types G and K. This sample includes the young stars AB~Dor, LQ~Hya, PZ~Tel, RX~J1508.6-4423 and the evolved stars HR~1099 and HD~199178. The only binary star of the sample is the primary component of the RS~CVn system HR~1099.

\subsection{General characteristics}

All studies of stellar differential rotation using parametric imaging or direct spot tracking (Cameron \& Donati 2002, Donati et al. 2003, Donati et al. 2000, Petit et al. 2003b) report detections of surface shears that are solar-like, i.e. stellar equator rotating faster than the polar regions. Early attempts at measuring the differential rotation of HR~1099, based on comparisons of Doppler images (Vogt et al. 1999, Strassmeier \& Bartus 2000), reported however an anti-solar shear for this star (the polar region rotating faster than the equator). A new study of this target, using the much more reliable method of parametric imaging, reports repeated detections of a solar differential rotation, both from brightness and magnetic tracers (petit et al. 2003a). 

It was suggested by Hackman et al. (2001) that in some cases, the flat bottom of the spectral line profiles of fast rotators, attributed by Doppler imaging codes to a high-latitude spot, may indeed be due to a distortion of line profiles under the effect of anti-solar differential rotation (an effect actually used by Reiners \& Schmitt 2002a to study differential rotation on unspotted stars). This possibility is already refuted in the cases of AB~Dor and HD~199178 (hosting a large polar spot) after the detection of a solar-like shear. Moreover, numerical simulations demonstrate that typical shear intensities observed on fast-rotating solar-type stars only generate a distortion of line profiles lower than $10^{-4}$ of the continuum level (distortions being stronger and potentially detectable, at a fixed shear strength d$\Omega$, for longer rotation periods), allowing us to safely assume that differential rotation does not produce significant biases in Doppler maps. 

All single stars of the sample have a shear intensity of the same order of magnitude, with beat periods (time for the equator to lap the pole by one rotation cycle) ranging from 40 to 110~d (120~d in the case of the Sun). The fact that beat periods of fast rotators are close to that measured on a slow rotator like the Sun suggests that the intensity of differential rotation is mostly independent of the rotation period. As a preliminary trend, it is also suggested that the shear intensity of young dwarfs is increasing with stellar mass (RX~J1508.6-4423, with the highest mass of the sample~-- 1.3$M_\odot$ --, has also the shortest beat period with 40~d only). This result nicely connects to recent shear measurements performed by Reiners \& Schmitt (2002) on more massive, slow rotating, unspotted F-type stars, reporting beat periods of the order of 10~d. However, they do not detect any mass dependence of the shear in their own sample, and further work from the same collaboration (Reiners \& Schmitt 2003) suggest that on such massive stars the shear strength may be significantly lower on fast rotators.

\subsection{The case of close binaries~: effect of tidal forces}

The surface shear measured on the primary component of the RS~CVn system HR~1099 (a K1 sub-giant) is significantly weaker than that derived from the sample of young single stars (with a beat period of HR~1099 of the order of 480~d, Petit et al 2003a). 

Owing to its evolutionary status, HR~1099 possesses a much deeper convective envelope than the young dwarfs observed before. To check whether stellar evolution was indeed partly responsible for this discrepancy, another evolved star was observed (the single FK~Com giant HD~199178, Petit et al 2003b). The beat period estimated in this last case is of the order of 90~days, a value very close to all measurements performed on main sequence stars. Another possibility is that the weaker differential rotation of HR~1099 is related to its lower mass (1~M$_\odot$, vs. 1.65~M$_\odot$ for HD~199178). However, the correlation between shear intensity and stellar mass was detected on young stars, and may not hold for more evolved objects. The effect that may most likely weaken the differential rotation in a close binary system like HR~1099 is the strong tidal forces that it suffers. As tides tend to synchronize rotation at all latitudes of the convective zone, only a weak shear is expected to survive (as predicted by Scharlemann 1981, 1982), in very good agreement with observations of HR~1099.

\subsection{Secular fluctuations of differential rotation}

\begin{figure}
\resizebox{\hsize}{!}
{\includegraphics[angle=270]{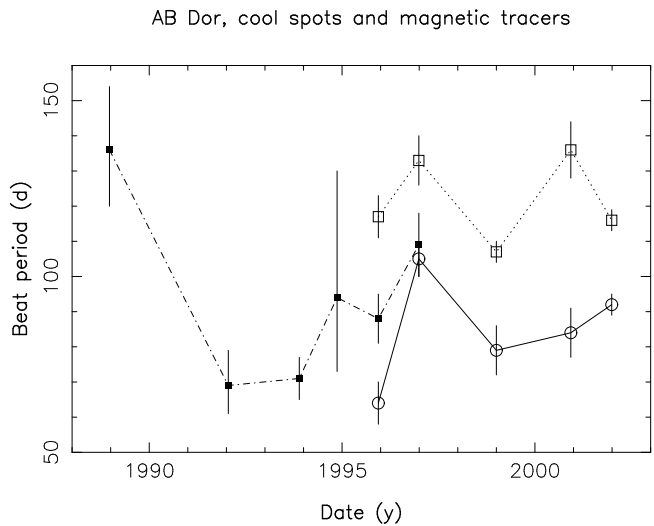}}
\caption{Differential rotation of AB~Dor as a function of time. The beat period is the time for the equator to lap the pole by one complete rotation cycle. Black squares represent measurements of Cameron \& Donati (2002), using a direct spot tracking method. White symbols represent estimates performed by means of parametric imaging (Donati et al. 2003). White squares correspond to cool spots, white circles to magnetic tracers.}
\label{abdor}
\end{figure}

Recent works by Cameron \& Donati (2002) and Donati et al. (2003) report temporal fluctuations in the differential rotation of the young dwarf AB~Dor, on timescales of several years (Fig. \ref{abdor}). The internal velocity field of the Sun itself is known to vary along the Solar magnetic cycle (Howard \& Labonte 1980, Howe et al. 2002, Vorontsov et al. 2002) but at a much lower level (variations of the rotation rate do not exceed 1~mrad.d$^{-1}$ in the Sun, whereas they can reach 40~mrad.d$^{-1}$ for AB~Dor). Similarly, the changes observed on AB~Dor may be related to its extremely active dynamo, and reveal the periodic conversion of kinetic energy into magnetic energy inside its convective zone (a mechanism proposed by Applegate 1992 to explain the orbital period fluctuations in close binary systems).

Another significant result, also obtained in the case of AB~Dor, is that the shear of magnetic regions is systematically stronger than that derived from cool spots (Fig. \ref{abdor}). This may indicate that these two types of tracers are not anchored at the same depth of the convective zone, and may therefore not be generated in the same layers, suggesting that the dynamo of AB~Dor may not be only active at the interface between its radiative core and its convective envelope (as opposite to the Sun), but presumably distributed within its whole convective zone. This conclusion is also suggested by the high level of shear fluctuations, hardly compatible with a large-scale magnetic field confined in a thin layer.

\section{Prospective}

The results detailed above, although very promising, can only be considered as a first step toward a more ambitious study of surface differential rotation of late-type stars. The preliminary trends already made out need to be confirmed on a larger sample of active stars, in order to determine how various physical mechanisms can impact the shear. Among the most promising stellar parameters to be investigated, we can mention the stellar mass, the evolutionary stage or the tidal forces at work in close binary systems. The detected temporal fluctuations of the shear also suggest that measurements of differential rotation must be regularly performed over several years to take into account the additional effect of long-term variability. Such a long-term monitoring should also enable one to get a more accurate insight on the feedback effect of stellar magnetic fields on the dynamics of convective zones along dynamo cycles.

The new generation of spectropolarimeters soon available (ESPaDOnS at the Canada-France-Hawaii Telescope, Hawaii, and NARVAL at the T\'elescope Bernard Lyot, France) will provide a significant increase in photon collecting power, spectral coverage and spectral resolution with respect to the current generation of instruments, therefore bearing the promise of major advances in the study of stellar differential rotation. Furthermore, the potential simultaneous use of ESPaDOnS and NARVAL to conduct dual-site observing campaigns, allowing an extremely dense phase sampling of targets, should also be a key improvement in the accuracy of measurements.


\acknowledgements

PP acknowledges the Portuguese {\em Funda\c c\~ao para a Ci\^encia e a Tecnologia} for grant support \#~SFRH/BPD/11139/2002.



\begin{thebibliography}{}
\bibitem{} Applegate, J.H.: 1992, ApJ 385, 621
\bibitem{} Barnes, J.R., Cameron, A.C., Unruh, Y.C, et al.: 1998, MNRAS 299, 904
\bibitem{} Barnes, J.R., Cameron, A.C., James, D.J., Donati, J.-F.: 2000, MNRAS 314, 162 
\bibitem{} Cameron, A.C., Robinson, R.D.: 1989, MNRAS 236, 57 
\bibitem{} Cameron, A.C., Donati, J.-F., Semel, M.: 2002, MNRAS 330, 699 
\bibitem{} Cameron, A.C., Donati, J.-F.: 2002, MNRAS 329, L23
\bibitem{} Cameron, A.C., Horne, K., Penny, A., Leigh, C.: 2002, MNRAS 330, 187
\bibitem{} Donati, J.-F., Brown, S.F.: 1997, A\&A 326, 1135
\bibitem{} Donati, J.-F., Semel, M., Carter, B.D., et al.: 1997, MNRAS 291, 658
\bibitem{} Donati, J.-F., Cameron, A.C.: 1997, MNRAS 291, 1  
\bibitem{} Donati, J.-F., Cameron, A.C., Hussain, G.A.J., Semel, M.: 1999, MNRAS 302, 437   
\bibitem{} Donati, J.-F., Mengel, M., Carter, B.D., Cameron, A.C., Wichmann, R.: 2000, MNRAS 316, 699  
\bibitem{} Donati, J.-F., Cameron, A.C., Petit, P.: 2003, MNRAS (in press)
\bibitem{} Hackman, T., Jetsu, L., Tuominen, I., 2001, A\&A 374, 171
\bibitem{} Howard, R., Labonte, B.J.: 1980, ApJL 239, L33
\bibitem{} Howe, R., Christensen-Dalsgaard, J., Hill, F., et al.: 2002, Sci 287, 2456
\bibitem{} Petit, P., Donati, J.-F., Cameron, A.C.: 2002, MNRAS 334, 374
\bibitem{} Petit, P., Donati, J.-F., Wade, G.A., et al.: 2003a, MNRAS (submitted)   
\bibitem{} Petit, P., Donati, J.-F., Wade, G.A., et al.: 2003b, MNRAS (submitted)   
\bibitem{} Reiners, A., Schmitt, J.H.M.M.: 2002, A\&A 393, L77
\bibitem{} Reiners, A., Schmitt, J.H.M.M.: 2003, A\&A (in press)
\bibitem{} Rice, J.B., Strassmeier, K.G.: 1996, A\&A 316, 164
\bibitem{} Scharlemann, E.T.: 1981, ApJ 246, 292
\bibitem{} Scharlemann, E.T.: 1982, ApJ 253, 298
\bibitem{} Strassmeier, K.G., Bartus, J.: 2000, A\&A 354, 537
\bibitem{} Vogt, S.S., Hatzes, A.P., Misch, A.A., K\"urster, M.: 1999, ApJS 121, 547 
\bibitem{} Vorontsov, S.V., Christensen-Dalsgaard, J., Schou, J., et al.: 2002, Sci 296, 101 
\end{thebibliography}
\end{document}